\documentclass[aps,preprint,nofootinbib,
superscriptaddress,showkeys]{revtex4-2}

\usepackage{color}
\usepackage{graphicx}
\usepackage{dcolumn}
\usepackage{bm}
\usepackage{commath}
\usepackage{amsmath}
\usepackage{upgreek}
\usepackage{siunitx}
\usepackage{physics}
\usepackage{natbib}
\usepackage[mathlines]{lineno}
\usepackage[english]{babel}
\usepackage[latin1]{inputenc}
\usepackage[bottom=2cm,top=2cm, left=1.5cm, right=1.5cm]{geometry}
\usepackage{booktabs}
\usepackage[unicode=true, bookmarks=true, bookmarksnumbered=false, bookmarksopen=false, breaklinks=false, pdfborder={0 0 1}, backref=false, colorlinks=false]{hyperref}
\usepackage[normalem]{ulem}
\begin{document}

\title{Distribution of antiferromagnetic rare-earth domains in multiferroic Dy$_{0.7}$Tb$_{0.3}$FeO$_3$}

\author{Yannik Zemp}
\affiliation{Department of Materials, ETH Zurich, Zurich, Switzerland}
\author{Ehsan Hassanpour}
\affiliation{UDEM Inselspital, University of Bern, Bern, Switzerland}
\author{Jan Gerrit Horstmann}
\affiliation{Department of Materials, ETH Zurich, Zurich, Switzerland}
\author{Yusuke Tokunaga}
\affiliation{Department of Advanced Materials Science, University of Tokyo, Chiba, Japan}
\author{Yasujiro Taguchi}
\affiliation{RIKEN Center for Emergent Matter Science (CEMS), Saitama, Japan}
\author{Yoshinori Tokura}
\affiliation{RIKEN Center for Emergent Matter Science (CEMS), Saitama, Japan}
\affiliation{Department of Applied Physics, University of Tokyo, Tokyo, Japan}
\author{Thomas Lottermoser}
\affiliation{Department of Materials, ETH Zurich, Zurich, Switzerland}
\author{Mads C. Weber}
\email{mads.weber@univ-lemans.fr}
\affiliation{Institut des Mol\'ecules et Mat\'eriaux du Mans, Le Mans Universit\'e, Le Mans, France}
\author{Manfred Fiebig}
\email{manfred.fiebig@mat.ethz.ch}
\affiliation{Department of Materials, ETH Zurich, Zurich, Switzerland}


\begin{abstract}
In many multiferroics, rare-earth and transition-metal orders exist side by side. For analyzing their interaction and its consequences for the multiferroic state, the associated domain patterns and their spatial correlation can give valuable insight. Unfortunately, this is often hampered by the lack of access to the domains of the rare-earth order. Here, we uncover such a domain pattern for the antiferromagnetic and multiferroic Dy$_{0.7}$Tb$_{0.3}$FeO$_3$. 
Optical second harmonic generation reveals the formation of column-like Dy/Tb domains. 
Interestingly, the columns form perpendicular to the magnetically induced electric polarization. 
Hence, the antiferromagnetic rare-earth order forces the ferroelectric domains to form nominally charged head-to-head and tail-to-tail domain walls, thus playing a leading role in the domain formation within the multiferroic phase. 
In turn, to reduce energy cost, the ferroelectric order causes a reduced rare-earth domain-wall density along the direction of the electric polarization. This interplay highlights the multiferroic character of the Dy$_{0.7}$Tb$_{0.3}$FeO$_3$ domain pattern.
We position Dy$_{0.7}$Tb$_{0.3}$FeO$_3$ within the broader landscape of rare-earth multiferroics and identify three distinct scenarios for the role of rare-earth order in these.

\end{abstract}

\maketitle

\section{Introduction}

Materials combining magnetic and ferroelectric order in a single phase are known as multiferroics. They are intensely studied because the ferroic coexistence can give rise to pronounced magnetoelectric coupling effects with the potential to control magnetic order by moderate electric voltage pulses rather than by energy-intensive current-generated magnetic fields. The first wave of multiferroic materials, including BiFeO$_3$, hexagonal YMnO$_3$, Ni$_3$B$_7$O$_{13}$I, or LiCoPO$_4$, was discovered in the 1960s~\citep{Schmid1994, Chupis2011}. They exhibit magnetic order of a single transition-metal sublattice, and the magnetic order occurs independently of the electric order, so that magnetoelectric coupling is not obvious.

In contrast, recent attempts are aimed at the development of multiferroics exhibiting pronounced magnetoelectric coupling, often in the form of the magnetic order inducing the electric one as improper ferroelectricity. Examples are orthorhombic TbMnO$_3$~\citep{Kimura2003a}, TbMn$_2$O$_5$~\citep{Hur2004}, Mn$_2$GeO$_4$~\citep{Honda2017}, or GdFeO$_3$~\citep{Tokunaga2009b}. Interestingly, many of these newer multiferroics (and also extensions of the ``historic'' multiferroic systems, like hexagonal ErMnO$_3$ as complement to YMnO$_3$) are oxides composed of both $3d$-transition-metal and $4f$-rare-earth ions. This raises the question of the participation of the rare-earth order in the manifestation of the multiferroic order. On the one hand, the multiferroic state in many systems emerges along with the magnetic (re-) order of the transition-metal ions between about 10 and 200~K, long before the rare-earth lattice orders. This is supported by the comparison of, for example, orthorhombic TbMnO$_3$ and YMnO$_3$, which, irrespective of the presence or absence of a magnetic rare-earth component, exhibit similar multiferroic phases~\citep{Mochizuki2011}. On the other hand, we have cases like orthorhombic TbMn$_2$O$_5$ and GdFeO$_3$, where the rare-earth order has undeniable influence on the multiferroic state~\citep{Hur2004, Lottermoser2009, Tokunaga2009b}.

A reliable way to identify correlations between ordered sublattices is to investigate the associated domain patterns. Coupling typically leads to correlations in the shape or distribution of domains, which are straightforward to recognize~\citep{Fiebig2023, Giraldo2021}.
However, there are very few investigations of this type because of the difficulty in accessing the rare-earth order with spatial resolution, in particular in the case of antiferromagnetism.

As a step in this direction, here we investigate the manifestation of the antiferromagnetic rare-earth domain pattern in multiferroic Dy$_{0.7}$Tb$_{0.3}$FeO$_3$. Using magnetoelectric order-parameter coupling in this material~\cite{Tokunaga2012,Hassanpour2022}, we derive the distribution of the rare-earth domains from the distribution of the ferroelectric domains that we image by spatially resolved optical second harmonic generation (SHG). We find highly anisotropic columnar Dy/Tb domains. 
Interestingly, the columnar shape the of the Dy/Tb domain does not appear to be dominated by the 4$f$-spin exchange interactions, as one would suspect, but potentially roots in soliton-lattice-type orders as found in DyFeO$_3$ and TbFeO$_3$.
Furthermore, the rare-earth domain pattern is not affected by the distribution or poling of the weakly ferromagnetic Fe domains that we access by Faraday-rotation microscopy (FRM). Hence, the formation of the Dy/Tb and Fe domains appears to occur independently. Nevertheless, the Dy/Tb and Fe lattices are exchange-coupled and establish the spontaneous electric polarization of Dy$_{0.7}$Tb$_{0.3}$FeO$_3$ and its distribution across the sample. One of the consequences is the formation of ferroelectric domain walls perpendicular to the direction of spontaneous polarization. Comparison of Dy$_{0.7}$Tb$_{0.3}$FeO$_3$ to other rare-earth multiferroics lets us distinguish three qualitatively different cases for the involvement of the rare-earth order in the multiferroic state.

\section{Experimental aspects} \label{setup}
The rare-earth orthoferrite Dy$_{0.7}$Tb$_{0.3}$FeO$_3$ crystallizes in a perovskite-type structure with the crystallographic space group \textit{Pbnm}~\citep{Geller1956, Geller1956a}. Below 653~K it exhibits antiferromagnetic \textit{G}-type order of the Fe lattice that is accompanied by a weak Dzyaloshinskii-Moriya-like magnetization~\citep{White1969,Tokunaga2012}. The antiferromagnetic spin alignment occurs along the $a$-axis, and the net magnetic moment is oriented along $c$ to reveal a \textit{G$_x$A$_y$F$_z$} order (in Bertaut's notation, with $x\,\|\,a$, $y\,\|\,b$, $z\,\|\,c$). In successive phase transitions, the Fe spins reorient to a \textit{F$_x$C$_y$G$_z$} order at 7~K, and back to the initial (\textit{G$_x$A$_y$F$_z$}) configuration at $\sim 2.65$~K. The latter transition is accompanied by purely antiferromagnetic \textit{G$_x$A$_y$}-type alignment of the Dy/Tb spins --- the order that is at the heart of this work~\citep{Tokunaga2012}. Together, the transition-metal and rare-earth orders induce an exchange-strictive electric polarization along the $c$-axis, making the material multiferroic with a magnetization $M_{\rm s}$ of 0.15~$\mu_{\rm B}$ per formula unit oriented parallel to a polarization $P_{\rm s}$ of 0.12~$\upmu$Ccm$^{-2}$ (space group \textit{Pb'n'2$_1$})~\cite{Tokunaga2009b,Tokunaga2012}. At 1.8~K, a final 90$^{\circ}$ spin reorientation of the Fe sublattice toward antiferromagnetic \textit{A$_x$G$_y$C$_z$} order occurs, while the \textit{G$_x$A$_y$}-type structure of the Dy/Tb sublattice is retained. This final magnetic phase transition leads to a breakdown of both the electric polarization and the net magnetization, now yielding $M_{\rm s}=P_{\rm s}=0$~\cite{Tokunaga2012,Hassanpour2021b}.

The symmetric exchange striction of the Fe--$3d$ and Dy/Tb--$4f$ spins driving the improper ferroelectric order can be parameterized by a free-energy contribution $F$ according to 
\begin{equation} \label{trilinear}
    F\propto -M\cdot L\cdot P,
\end{equation}
where $M$, $L$, and $P$ represent the magnetic transition-metal order, the purely antiferromagnetic rare-earth order, and the improper electric polarization, respectively~\cite{Tokunaga2009b,Tokunaga2012,Hassanpour2022}. With $M$, $L$, and $P$ each normalized to $\pm 1$, we thus get $M_0L_0P_0\equiv -1$. It is a remarkable consequence of this trilinear coupling that despite the rigidity of this relation, $M$, $L$, and $P$ can each develop different domain structures as long as the trilinear product is conserved. Hence, if by poling experiments one of the three ordered states is transferred into a single-domain configuration, the other two are forced to exhibit identical domain patterns. To support this, it was demonstrated in a series of poling experiments, how to act selectively on $M$, $L$, and $P$ with appropriately timed magnetic- or electric-field pulses~\citep{Hassanpour2022}.

Here we exploit the trilinear correlation to image the distribution of antiferromagnetic rare-earth domains, which would normally be out of reach because of the absence of a magnetization to couple to. Specifically, we transfer the weakly ferromagnetic Fe system into a single-domain configuration by magnetic-field cooling. Using optical SHG, we image the ferroelectric domain pattern, which must now be identical to the Dy/Tb domain pattern. In other words, we detect the domain distribution of $P$ with $P\propto ML$ according to Eq.~(\ref{trilinear}), and even $P\propto L$ when $M$ is uniform (single-domain).

\begin{figure*}[h!]
\centering
\includegraphics[width=0.75\columnwidth]{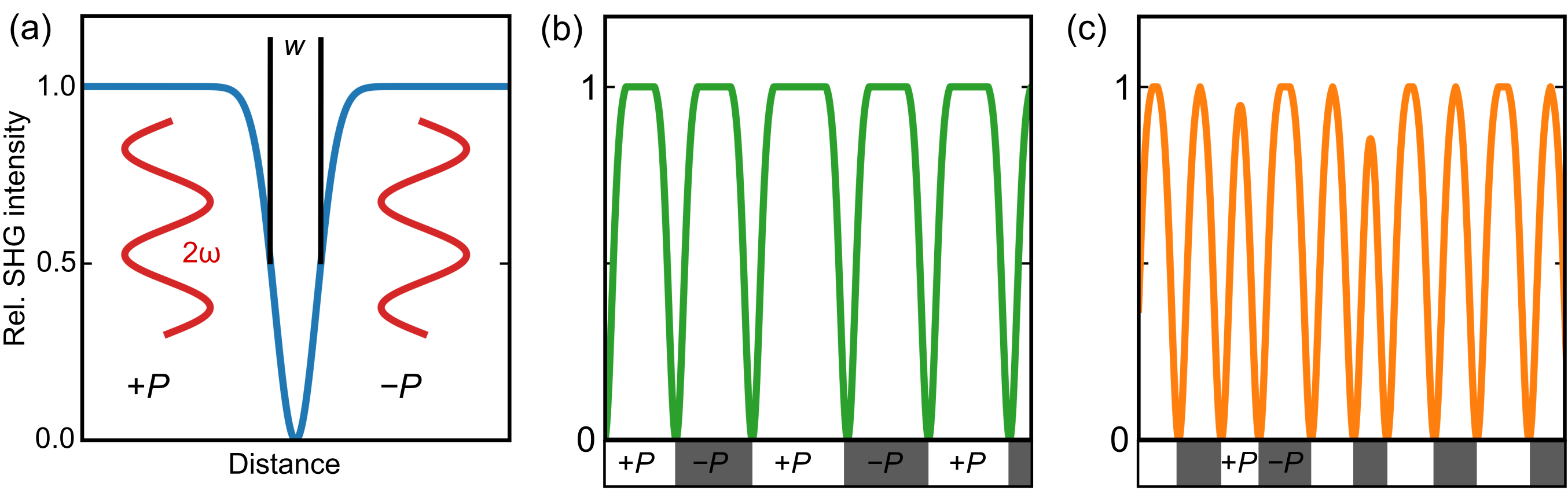}
\caption{Destructive SHG interference at domain walls. (a) SHG intensity dependence for a ferroelectric sample with a single $\pm P$ domain wall. At the wall, the SHG light fields from the neighboring domains interfere destructively so that the region at the wall appears as dark groove with the width $w$ of the optical spatial resolution. (b) SHG intensity dependence for multiple ferroelectric domains with a width~$\gtrsim w$. (c) SHG intensity dependence for multiple ferroelectric domains with a width~$\lesssim w$.}
\label{dw-image} 
\end{figure*}

SHG denotes the frequency doubling of a light wave in a material and is described by the relation $P_i(2\omega)=\epsilon_0\chi_{ijk}E_j(\omega)E_k(\omega)$~\cite{Fiebig2023,Denev2011,shen2003principles}. Here, $\vec{E}(\omega)$ is the electric field of the incident fundamental light, and $\vec{P}(2\omega)$ is the frequency-doubled polarization generated in the material, where the latter is the source of the SHG light emitted from the crystal. In addition, $\epsilon_0$ is the vacuum permittivity, and $\hat{\chi}$ is the nonlinear susceptibility tensor. SHG is sensitive to breaking of the inversion symmetry by ferroelectric polarization $P$. It produces a light wave with a phase that is proportional to the sign of $P$ because of $\hat{\chi}\propto P$. Hence, a sign change of $P$ leads to a sign change of the SHG wave and causes destructive interference in the vicinity of the $\pm P$ domain walls. This highlights the ferroelectric domain walls as dark grooves and thus enables mapping of the ferroelectric domain pattern~\cite{Fiebig2023, Denev2011}. The concept of this process is sketched in Fig.~\ref{dw-image}a. The SHG polarization components that can be nonzero are determined by $\hat{\chi}$ and thus by the symmetry of the multiferroic phase. Specifically, the \textit{m'm'2} point-group symmetry of this phase in Dy$_{0.7}$Tb$_{0.3}$FeO$_3$ leads to the independent susceptibility components $\chi_{aac}=\chi_{aca}$, $\chi_{bbc}=\chi_{bcb}$, $\chi_{caa}$, $\chi_{cbb}$, $\chi_{ccc}$, $\chi_{abc}=\chi_{acb}$, $\chi_{bca}=\chi_{bac}$, $\chi_{cab}=\chi_{cba}$~\cite{Birss1966a}. Extended measurements revealed that the $\chi_{ccc}$ component leads to the highest SHG yield, so we used it for all SHG imaging experiments.

Dy$_{0.7}$Tb$_{0.3}$FeO$_3$ single crystals were grown by the floating-zone technique and cut into platelets of $3\times 3\times 1$~mm$^3$ oriented perpendicular to the $a$-, $b$-, or $c$-axis. They were thinned down to $60-90$~$\upmu$m by lapping with Al$_2$O$_3$ powder and chemo-mechanically polished on both sides with silica slurry to achieve a surface roughness of $\sim 1$~nm. The samples were mounted in an optical liquid-helium-cooled cryostat generating magnetic fields up to 10~T at temperatures of 1.6 -- 325~K (Oxford Instruments, Spectromag). The samples were probed with light from an amplified Ti:sapphire laser system generating pulses at 1.55~eV (800~nm) at a repetition rate of 1~kHz and with a pulse length of 120~fs. The laser light was sent into an optical parametric amplifier to generate radiation with a photon energy of $0.5-2.1$~eV ($2700-600$~nm) and a pulse energy of 1-200~\textmu J (Coherent OPerA Solo). The Dy$_{0.7}$Tb$_{0.3}$FeO$_3$ samples were excited in a transmission setup described elsewhere~\citep{Fiebig2023}, and SHG light was detected with spatial resolution using a telephotography lens (Soligor 135~mm, $1:3.5$) and a liquid-nitrogen-cooled digital camera (Photometrics AT200). For imaging experiments, we chose an SHG photon energy of 1.9~eV where the $\chi_{ccc}$-related SHG spectrum of Dy$_{0.7}$Tb$_{0.3}$FeO$_3$ exhibits a maximum.

In contrast, measurements by FRM were used to probe the distribution of the weakly ferromagnetic Fe order. Such experiments were restricted to a $c$-oriented sample because of $M\,\|\,c$. For FRM image capture, we utilized a light-emitting diode that emits at a central wavelength of 660\,nm (Thorlabs M660L4), paired with a monochromatic digital camera (The Imaging Source DMK 22BUC03).

\section{Experimental results}

\begin{figure*}[h!] \centering
\includegraphics[width=0.5\columnwidth]{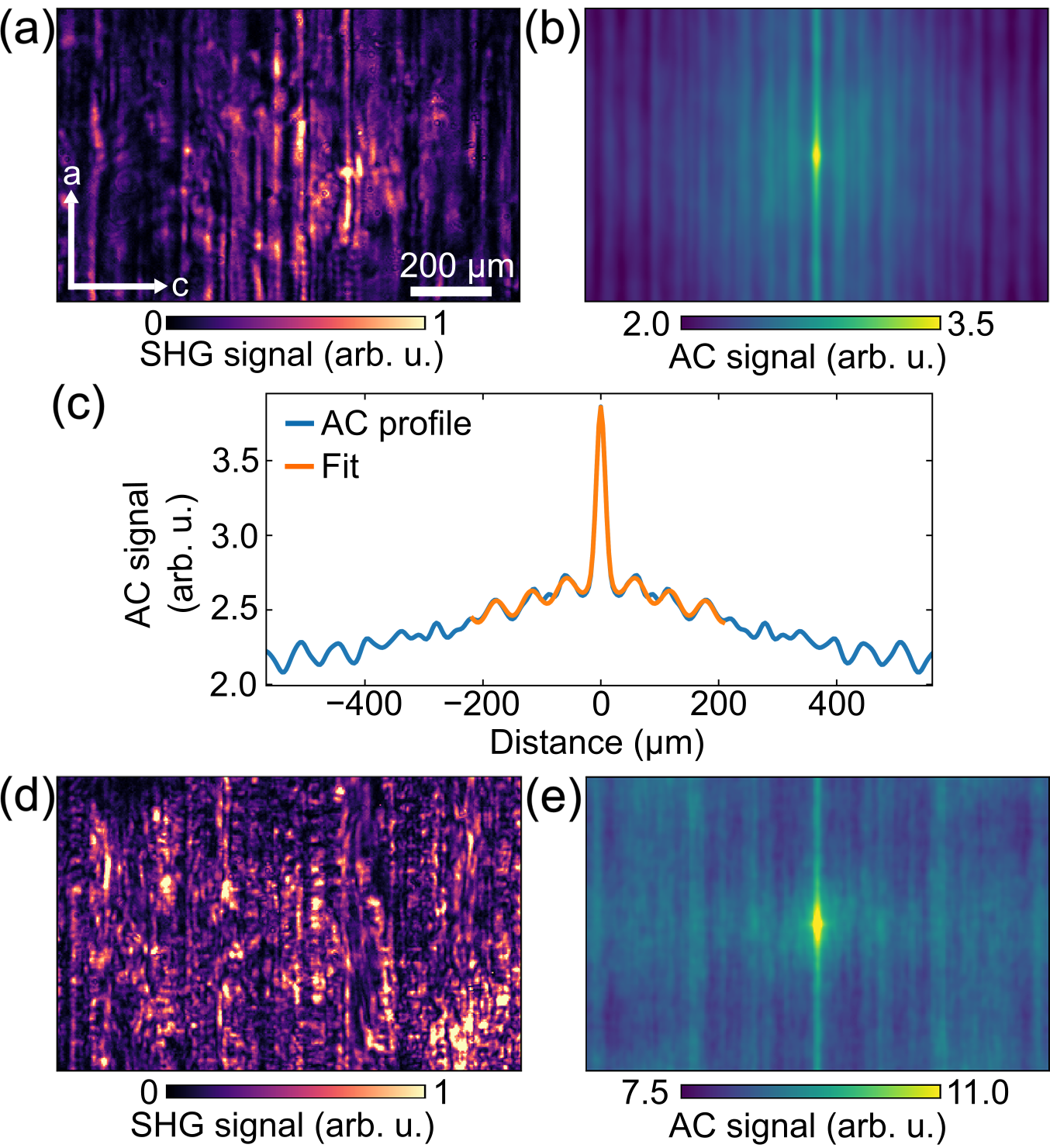}
\caption{Spatially resolved distribution of SHG intensity on a $b$-oriented Dy$_{0.7}$Tb$_{0.3}$FeO$_3$ sample after magnetic-field cooling. (a) The SHG signal images the ferroelectric domain pattern in $P$, which is, according to Eq.~(\ref{trilinear}), identical to the antiferromagnetic rare-earth domain pattern in $L$ because the weakly-ferromagnetic Fe lattice $M$ is in a single-domain configuration. The sample exhibits stripe-like rare-earth domains along the $a$-direction. (b) Autocorrelation (AC) function of the image in (a). Periodic lines identify the average width of the stripe-like domains. (c) Cross section of the autocorrelation signal in (b). The orange solid line represents a fit of two Gaussians and a sine-like modulation. (d) Spatially resolved distribution of SHG intensity on a $b$-oriented Dy$_{0.7}$Tb$_{0.3}$FeO$_3$ sample after zero-field cooling. Because here the Fe lattice is in a multi-domain configuration, the domain pattern in $P$ now represents a convolution of $M$- and $L$-domains. Visually, this leads to a superposition of the granular $M$-domains in Fig.~\ref{frm-image}b and the stripe-like $L$-domains in Fig.~\ref{b-image}a. (e) Autocorrelation function of the image in (d). Lines along the $a$-direction clearly identify the stripe-like $L$ domains.}
\label{b-image}
\end{figure*}

Figure~\ref{b-image}a shows the spatially resolved distribution of SHG light on a $b$-oriented Dy$_{0.7}$Tb$_{0.3}$FeO$_3$ sample that was cooled to the multiferroic phase in a magnetic field $\mu_0H_c=300$~mT, where $\mu_0$ is the vacuum permeability. The purpose of the field cooling was to convert the sample into a weakly ferromagnetic Fe single-domain configuration. As explained in Section \ref{setup} [Eq.~(\ref{trilinear})], the ferroelectric domain pattern reproduced by the SHG image in Fig.~\ref{b-image}a must then be identical to the distribution of the antiferromagnetic rare-earth domains. Figure~\ref{b-image}a reveals highly anisotropic domains that extend across the entire length of the image in the $a$-direction, but are closely stacked along the $c$-axis, reminiscent of the sketch in Fig.~\ref{dw-image}b. Figure~\ref{b-image}b shows the autocorrelation function of the image in panel (a), and a horizontal line scan across panel (b) is depicted in Figure~\ref{b-image}c. The autocorrelation function is obtained by taking the inverse Fourier transform of the power spectrum, which itself is the absolute square of the Fourier transform of the image~\cite{Robertson2012}. It reveals a periodic distribution of lines that can be fitted with two Gaussians, capturing the central peak and the intensity distribution away from it, and a sine-function reproducing the intensity oscillation. The period of this oscillation represents the average width of the stripe domains in Fig.~\ref{b-image}a, for which the fit gives a value of $59.4\pm 0.2$~$\upmu$m.

Before we look at the distribution of the antiferromagnetic rare-earth domains in other crystal directions, it should be excluded that the rare-earth domain structure observed in Fig.~\ref{b-image}a is influenced in any way by the weakly ferromagnetic Fe order. We know that for a single magnetic Fe domain, the trilinear coupling [Eq.~(\ref{trilinear})] provides the freedom for the rare-earth domain pattern to be organized in any preferred way. In Fig.~\ref{b-image}d, we also investigate the case of a multi-domain configuration of the Fe order. The SHG image in Fig.~\ref{b-image}d was obtained in the same way as Fig.~\ref{b-image}a, but now on a zero-field-cooled $b$-oriented Dy$_{0.7}$Tb$_{0.3}$FeO$_3$ sample. Compared to the stripe-like distribution of SHG intensity in Fig.~\ref{b-image}a, we observe a predominantly granular structure in Fig.~\ref{b-image}d. According to the discussion of Eq.~(\ref{trilinear}) in Section \ref{setup}, the grainy structure can result from a change in the distribution of the antiferromagnetic rare-earth domains ($\sim L$) or from a multi-domain configuration of the weakly ferromagnetic Fe domains ($\sim M$) in the zero-field-cooled sample, since SHG images the distribution of the ferroelectric domains with $P\propto ML$. Two arguments support the grainy SHG distribution as a result of the weakly ferromagnetic multi-domain configuration in $M$, while the antiferromagnetic $L$-stripe-domain configuration remains unaffected by the change from field cooling to zero-field cooling.

\begin{figure*}[h!] 
\centering
\includegraphics[width=0.5\columnwidth]{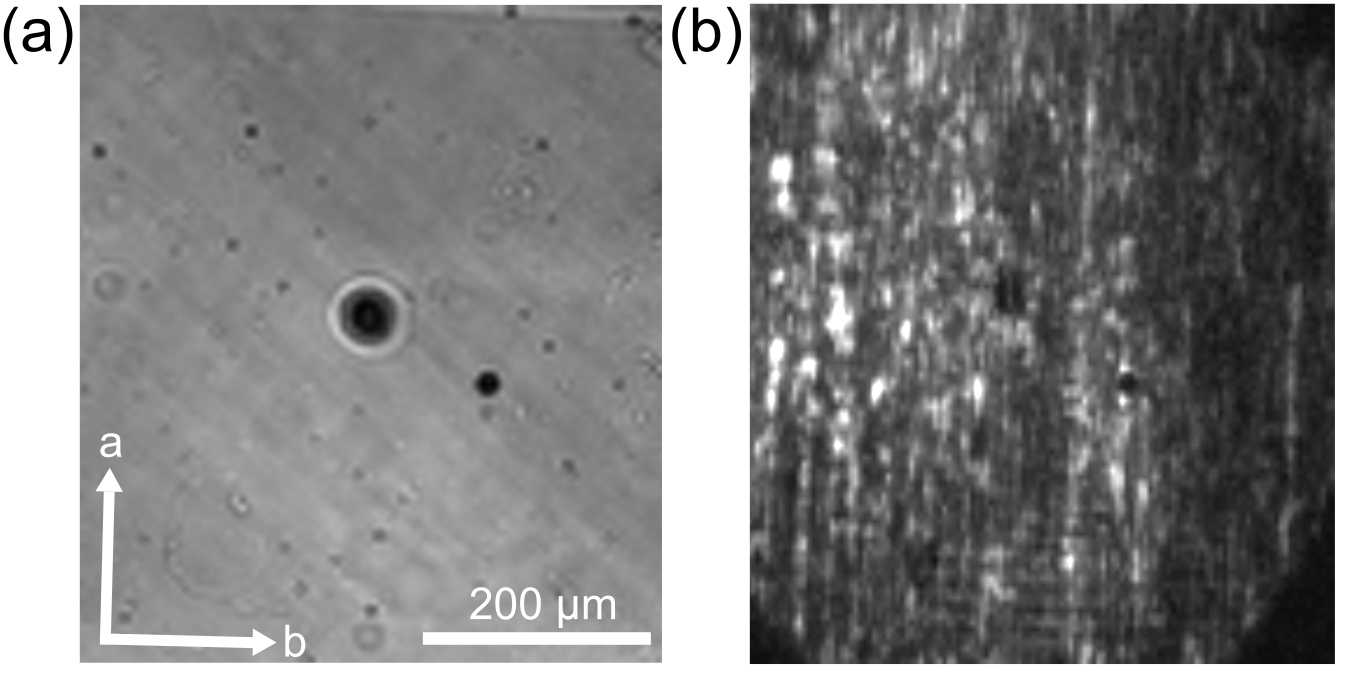}
\caption{FRM image of the distribution of weakly ferromagnetic Fe domains obtained on a $c$-oriented Dy$_{0.7}$Tb$_{0.3}$FeO$_3$ sample. (a) After magnetic-field-cooling the homogeneous distribution of intensity points to a single-domain configuration of the sample in $M$ (the dark center spot is a small hole in the sample). (b) After zero-field-cooling, the granular stripe-like distribution of intensity shows a multi-domain configuration of the sample in $M$. Since FRM integrates the signal along the light propagation direction, the measurement effectively probes the entire three-dimensional state of the sample, with varying contrast indicating a multidomain state across all three dimensions. This allows for a meaningful comparison of the FRM results with the SHG results, even for the differently oriented samples shown in Figs.~\ref{b-image} and \ref{a-image}.}
\label{frm-image}
\end{figure*}

First, the complementary FRM image in Fig.~\ref{frm-image}b yields a grainy intensity distribution similar to that of the SHG image. It strikingly contrasts with the homogeneous distribution after field cooling in Fig.~\ref{frm-image}a and shows that the weakly ferromagnetic Fe sublattice has indeed been transferred to a configuration of multiple domains close to the optical resolution limit. Second, despite the convolution with this multi-domain configuration, the SHG image still reveals the stripe-like intensity distribution that is characteristic of the rare-earth domain pattern after field cooling. This indicates that the stripe-like distribution of antiferromagnetic rare-earth domains prevails irrespective of the multi-domain configuration of the weakly ferromagnetic Fe order. 
We therefore conclude that the intrinsic antiferromagnetic rare-earth domain pattern is not affected by the Fe domain order.

\begin{figure*}[h!] 
\centering
\includegraphics[width=0.5\columnwidth]{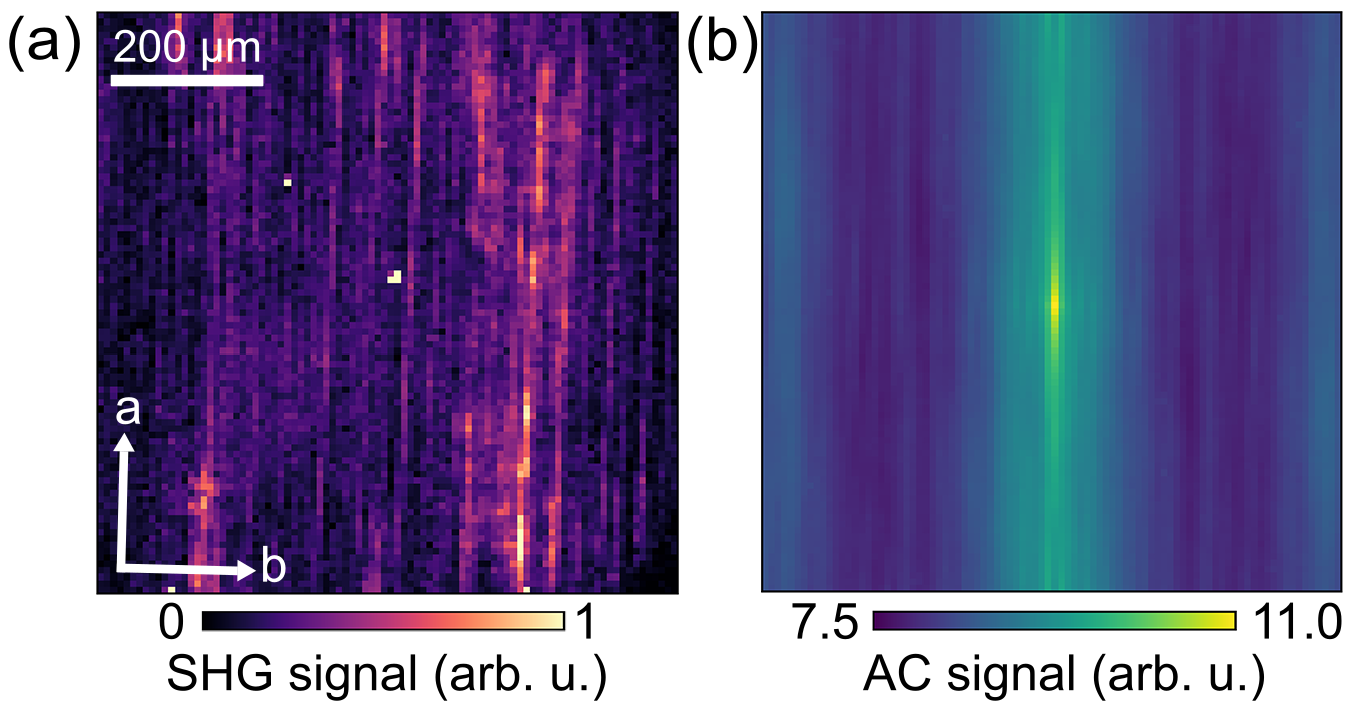}
\caption{Spatially resolved distribution of SHG intensity on a $c$-oriented Dy$_{0.7}$Tb$_{0.3}$FeO$_3$ sample after magnetic-field cooling. (a) The SHG signal images the ferroelectric domain pattern in $P$, which is identical to the antiferromagnetic rare-earth domain pattern in $L$ because the weakly-ferromagnetic Fe lattice $M$ is in a single-domain configuration. The sample exhibits stripe-like domains along the $a$-direction. (b) Autocorrelation (AC) function of the image in (a). The lines corresponding to the stripe-like domains do not exhibit a clear periodicity which points to an average domain width slightly below the optical resolution of 15~$\upmu$m.}
\label{c-image}
\end{figure*}

With this background knowledge, we can now analyze the distribution of antiferromagnetic rare-earth domains in the $a$- and $c$-oriented samples, now again after magnetic-field cooling so that the ferroelectric domain pattern obtained by SHG reproduces the antiferromagnetic rare-earth domain pattern. Figure~\ref{c-image} shows the spatially resolved distribution of SHG light on a $c$-oriented Dy$_{0.7}$Tb$_{0.3}$FeO$_3$ sample and the associated autocorrelation function. Note that to access $\chi_{ccc}$ on this sample, we tilted the sample by $36^{\circ}$ around the $a$-axis. Just as in Fig.~\ref{b-image}, we find a distribution of domain stripes running along the $a$-axis. The density of the stripes is much higher than in the case of the $b$-oriented sample, however. In fact, the autocorrelation function no longer reveals the intensity oscillations associated with the width of these stripes. Apparently, this width is at or below the optical resolution of our experiment and, hence, $< 15$~$\upmu$m. The SHG image in Fig.~\ref{c-image}a therefore resembles the situation sketched in Fig.~\ref{dw-image}c.

\begin{figure*}[h!] \centering
\includegraphics[width=0.5\columnwidth]{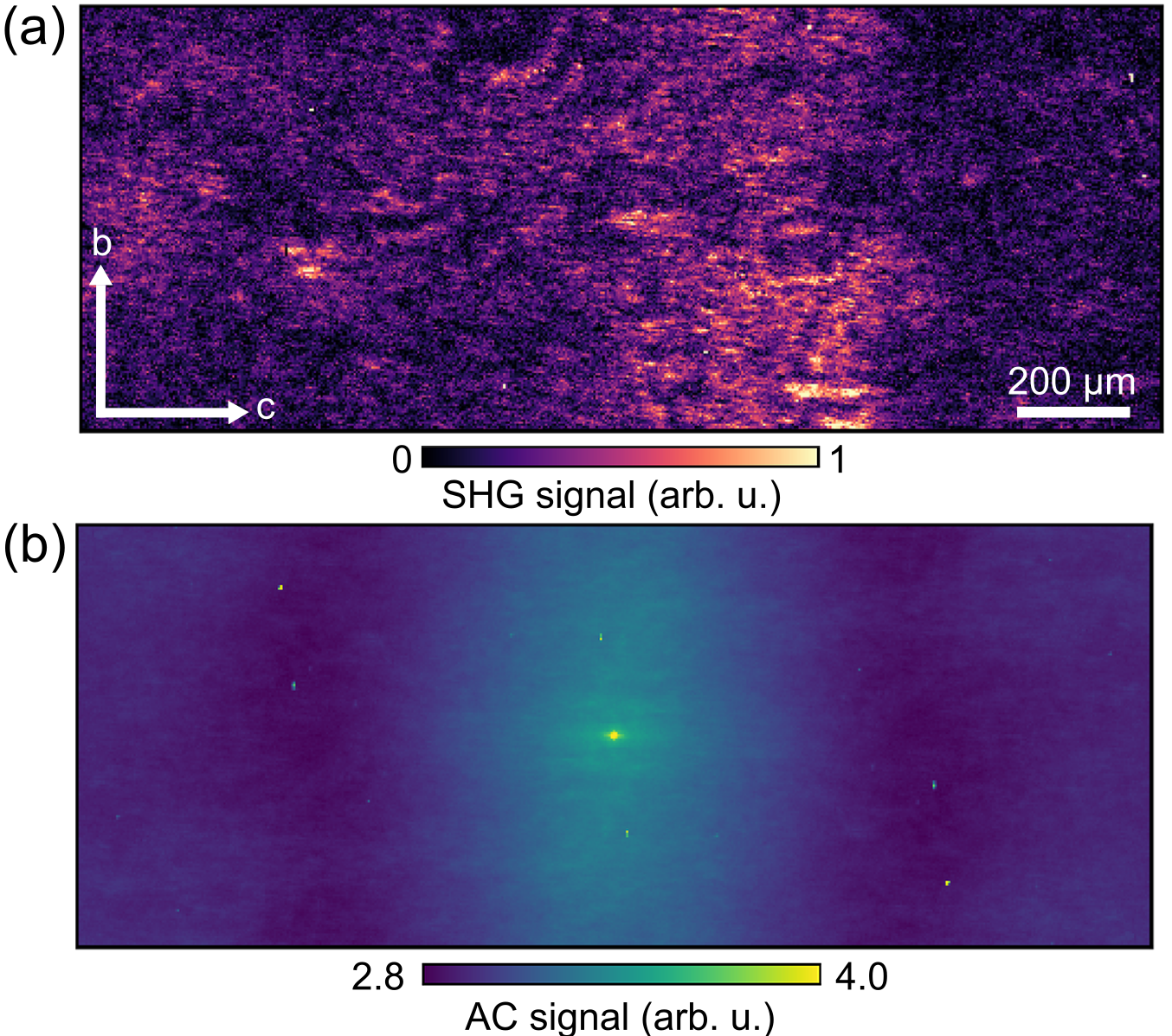}
\caption{Spatially resolved distribution of SHG intensity on a $a$-oriented Dy$_{0.7}$Tb$_{0.3}$FeO$_3$ sample after magnetic-field cooling. (a) The SHG signal images the ferroelectric domain pattern in $P$, which is identical to the antiferromagnetic rare-earth domain pattern in $L$ because the weakly-ferromagnetic Fe lattice $M$ is in a single-domain configuration. The granular distribution of SHG intensity reveals an extension of the domains in $b$- and $c$-directions that is close to the optical resolution limit. (b) Autocorrelation (AC) function of the image in (a). In confirmation of the granular structure in (a), lines that would indicate a presence of stripe-like domains are no longer visible.}
\label{a-image}
\end{figure*}

Finally, Fig.~\ref{a-image} shows the spatially resolved distribution of SHG light on an $a$-oriented Dy$_{0.7}$Tb$_{0.3}$FeO$_3$ sample, again accompanied by the autocorrelation function. We now find a granular distribution of SHG intensity without any stripes. This indicates a dense distribution of domains with an extension on the order of the optical resolution in both the $b$- and $c$-directions. 

Combining the results of the imaging experiments in all three directions, we conclude that the stripes visible in Figs.~\ref{b-image} and \ref{c-image} represent columnar domains whose long axis is directed along the $a$-direction. The base plate of the columns is elliptical with an average extension of $\sim 60$~$\upmu$m in the $c$-direction and $< 15$~$\upmu$m in the $b$-direction. In Fig.~\ref{a-image}, we see the top of these columns. The visual impression of an elongation of the granular structure in Fig.~\ref{a-image} in the $c$-direction seems to be consistent with the anisotropic cross-section of the domain columns, but is not visible in the autocorrelation function.

\section{Concluding discussion}

Using the trilinear coupling of the weakly ferromagnetic order of the Fe lattice, the antiferromagnetic order of the Dy/Tb lattice, and the ferroelectric order induced by $3d$-$4f$ spin exchange striction, we image the distribution of the rare-earth antiferromagnetic domains in multiferroic Dy$_{0.7}$Tb$_{0.3}$FeO$_3$ in spatially resolved SHG experiments. We find a distribution of columnar domains with an elliptical base. The extension of the domains is $\gtrsim 1$~mm along the $a$-axis, $< 15$~$\upmu$m along the $b$-axis, and $\sim 60$~$\upmu$m along the $c$-axis. Despite the pronounced magnetoelectric coupling, the shape and distribution of the rare-earth domains do not appear to be affected by the domain configuration of the transition-metal sublattice. In contrast, there are signs that the domain configuration of the rare-earth lattice influences that of the transition-metal lattice rather than the other way around. Specifically, the FRM image at 2~K in Fig.~\ref{frm-image}b reveals traces of Fe domain stripes along the $a$-axis that are reminiscent of the Dy/Tb domain stripes. These stripe-like features in the Fe domain pattern contrast with the typical bubble- or maze-like domain pattern of the Fe order exhibits above 7~K~\cite{HassanpourYesaghi2019}. 
Yet a conclusive statement whether the stripe-like traces of the Fe domains at 2\,K result from an interaction with the rare-earth order or are intrinsic to the Fe order requires further investigation. 

\begin{figure*}[h!] \centering
\includegraphics[width=0.75\columnwidth]{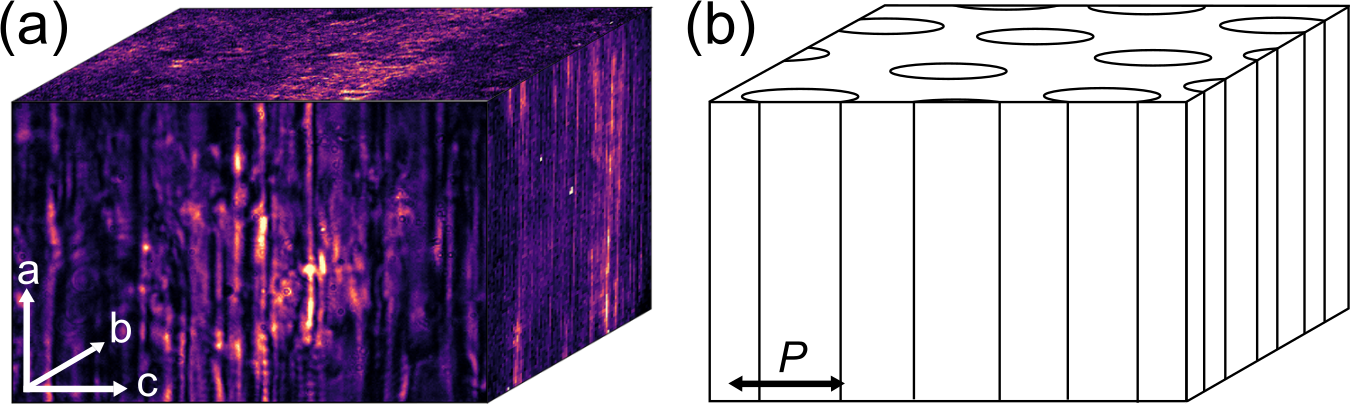}
\caption{(a) Three-dimensional visualization of the rare-earth domain patterns of Figs.~\ref{b-image}a, \ref{c-image}a, and \ref{a-image}a. (b) Conceptual sketch of the shape and distribution of the rare-earth domains in Dy$_{0.7}$Tb$_{0.3}$FeO$_3$. $\pm P$ indicates the direction of the magnetically induced polarization perpendicular to the column-shaped domains resulting in nominally charged domain walls.}
\label{3d-image}
\end{figure*}

Because ferroelectric polarization is improper, its domain pattern is determined by the domain patterns of the magnetic sublattices. Hence, in agreement with the secondary nature of the spontaneous polarization, the ferroelectric domains are forced to exhibit the same columnar domain pattern as the antiferromagnetic Dy/Tb sublattice. 
Since the columns are oriented perpendicular to the spontaneous electric polarization, the ferroelectric domain pattern exhibits nominally charged head-to-head and tail-to-tail domain walls in the $ac$-plane, which a proper-ferroelectric material would avoid. Yet to limit the energy expense that comes with charged domain walls, it is favorable to reduce the domain wall density in the $ac$-plane compared to the $ab$-plane. This is a likely cause of the elliptical shape of the domains (see Fig.~\ref{3d-image}).
Hence, the domain pattern appears to be dominated by the antiferromagnetic rare-earth order that dictates the formation of strongly anisotropic columns. Their elliptical shape however, would be driven by the ferroelectric order to minimize the number of charged domain walls. The domain pattern therefore exhibits a multiferroic character with a clear hierarchy.

To understand the origin of the pronounced anisotropy in the Dy/Tb-domain pattern itself, detailed knowledge of the microscopic Dy/Tb-4$f$-spin interactions is crucial. 
However, this information is currently missing due to the lack of neutron diffraction data for Dy$_{0.7}$Tb$_{0.3}$FeO$_3$. 
For Dy, the dominant rare-earth ion in Dy$_{0.7}$Tb$_{0.3}$FeO$_3$, comparisons with other Dy-based ortho-perovskites, such as DyFeO$_3$, DyScO$_3$ and DyAlO$_3$, suggest that the 4$f$-spin exchange interaction along the $c$-axis dominates~\cite{Wu2017, Gorodetsky1968a, Schuchert1969}. 
This would indicate that the domain walls aligned with the $c$-axis as the primary spin interaction direction~\cite{ Moromizato2024}. Yet, in Dy$_{0.7}$Tb$_{0.3}$FeO$_3$, the observed domain walls align with the $a$-axis, coinciding with the $G_x$ spin order. This deviation points to a more complex interplay of spin interactions that cannot be captured by this approach alone. As a possibility, the domain pattern might be explained by a soliton lattice characteristic intrinsic to its parent compounds~\cite{Artyukhin2012,Nikitin2025}. However, this would require additional neutron scattering experiments which goes beyond the scope of this work.

Our work highlights the crucial role of the rare-earth order in the formation of the multiferroic order of Dy$_{0.7}$Tb$_{0.3}$FeO$_3$. The Dy/Tb order is relevant not only for the emergence of the multiferroic state through exchange striction with the Fe transition-metal lattice but also it is particularly significant for domain formation in the multiferroic phase. 
Although the magnetic rare-earth order directly contributes to the emergence of multiferroicity in this work, future studies may include systems in which the relations between the rare-earth and transition-metal sublattices are different from Dy$_{0.7}$Tb$_{0.3}$FeO$_3$. 
In total, we can discern three different cases with different roles of the rare-earth ions in the multiferroic phases. On all of these the knowledge about the rare-earth domain pattern would allow an insight into potential magnetoelectric interactions even if they are subtle. 

(i)	Materials like Dy$_{0.7}$Tb$_{0.3}$FeO$_3$, where the rare-earth and transition-metal sublattices are jointly responsible for the emergence of the improper electric polarization. Here, the rare-earth ions intrinsically contribute in a significant way to the multiferroic domain formation, reflecting the strong magnetoelectric coupling and the essential role of the rare earth therein. 

(ii + iii) Materials that are already multiferroic in the absence of magnetic rare-earth order so that the latter merely modifies an existing multiferroic state. 
Here we distinguish (ii) a case like orthorhombic TbMnO$_3$, where the electric polarization is magnetically induced~\citep{Katsura2005, Mostovoy2006} so that the influence of the rare-earth system in this case is likely because of the inherent sensitivity of the polar order to changes in the magnetic environment. In TbMn$_2$O$_5$, macroscopic measurements suggest the influence of the rare-earth order on the polarization even more distinctly~\citep{Leo2012,Lottermoser2009}. Access to the rare-earth domain pattern could give insight as to the extent to which this inherent sensitivity manifests itself in magnetoelectric interaction with the rare-earth order. We also have (iii) a case such as hexagonal ErMnO$_3$, where the ferroelectric order emerges independently of the magnetic order. The Mn and rare-earth orders are characterized by strong crosstalk~\citep{Meier2012a}. However, it is less obvious if the polar order~\citep{Fennie2005, Lilienblum2015a, Tosic2022} also influences the rare-earth system or even vice versa.
A crucial question is whether the interaction between the rare-earth and transition-metal ions already defines the domain pattern of the rare-earth element.
In this context, it has been shown in the non-multiferroic TmFeO$_3$ that a magnetic order of the rare-earth ion induced by the magnetic Fe order can form a domain pattern independent of the Fe domain configuration~\citep{Staub2017}. 
This shows that even such an induced rare-earth domain pattern can have an independent domain configuration. 

\newpage

\section*{Acknowledgements}

This work was financially supported by the Swiss National Science Foundation (SNSF) via projects No.\ 200021{\_}178825, 200021{\_}215423, TMPFP2\_217303. MCW is grateful for financial support by the R\'egion des Pays de la Loire under the Etoiles Montantes Initiative (2022{\_}11808) and the PULSAR Academy (2022{\_}09767). This work was supported by a government grant of France managed by the ANR as a part of the France 2030 investment plan from PEPR SPIN ANR-22-EXSP 0003. The authors thank Quintin N. Meier for fruitful discussions.

\bibliography{library.bib}

\end{document}